\documentstyle[12pt,epsfig,amsfonts,latexsym]{article}

\makeatletter

\@addtoreset{equation}{section} \makeatother

\newcommand{\be}{\begin{equation}}

\newcommand{\ee}{\end{equation}}

\newcommand{\ba}{\begin{eqnarray}}

\newcommand{\ea}{\end{eqnarray}}

\newcommand{\nn}{\nonumber}

\newcommand{\bea}{\begin{eqnarray*}}

\newcommand{\eea}{\end{eqnarray*}}

\newcommand{\gsim}{\raise.3ex\hbox{$>$\kern-.75em\lower1ex\hbox{$\sim$}}}

\newcommand{\lsim}{\raise.3ex\hbox{$<$\kern-.75em\lower1ex\hbox{$\sim$}}}

\begin{document}

\titlepage

\begin{flushright}
\today \\
hep-th/0411058 \\
\end{flushright}
\vskip 1cm
\begin{center}
{\large \bf  Effective actions  of a Gauss-Bonnet brane
world with brane curvature terms}
\end{center}

\vspace*{5mm} \noindent

\centerline{Ph.~Brax\footnote{brax@spht.saclay.cea.fr}${}^{a}$,
N.~Chatillon\footnote{chatillon@spht.saclay.cea.fr}${}^{a}$,
D.A.~Steer\footnote{steer@th.u-psud.fr}${}^{b}$}

\vskip 0.5cm \centerline{${a}$) \em Service de Physique
Th\'eorique} \centerline{\em CEA/DSM/SPhT, Unit\'e de recherche
associ\'ee au CNRS,} \centerline{\em CEA-Saclay F-91191 Gif/Yvette
cedex, France.} \vskip 0.5cm

\centerline{ $b$) \em Laboratoire de Physique
Th\'eorique\footnote{Unit\'e Mixte de Recherche du CNRS (UMR
8627)}, B\^at. 210,} \centerline{\em Universit\'e Paris XI, 91405
Orsay Cedex, France}  \centerline{\em and } \centerline{\em
F\'ed\'eration de recherche APC, Universit\'e Paris VII, 2 place
Jussieu,} \centerline{\em 75251 Paris Cedex 05, France.} \vskip
2cm

\begin{center}
{\bf Abstract}
\end{center}

We consider a warped brane world scenario with two branes,
Gauss-Bonnet gravity in the bulk, and brane localised curvature
terms. When matter is present on both branes, we investigate the
linear equations of motion and distinguish three regimes. At very
high energy and for an observer on the positive tension brane,
gravity is four dimensional and coupled to the brane bending mode
in a Brans-Dicke fashion. The coupling to matter and brane bending
on the negative tension brane is exponentially suppressed. In an
intermediate regime, gravity appears to be five dimensional while
the brane bending mode remains four dimensional. At low energy,
matter on both branes couple to gravity for an observer on the
positive tension brane, with a Brans-Dicke description
similar to the 2--brane Randall-Sundrum setup. We
also consider the zero mode truncation at low energy and show that
the moduli approximation fails to reproduce the low energy action.

\newpage
\section{Introduction}
Since the pioneering work of Randall and Sundrum \cite{RS1,RS2},
brane world models have been studied intensively.  In the simplest
setup, which provides a potential solution of the hierarchy
problem, two branes of tension $T_i$ ($i=\pm$) are embedded in a
5D bulk AdS spacetime with negative cosmological constant
$\Lambda$. The action for the system is
\begin{equation}
S_{{\rm RS}} = \frac{1}{2 \kappa_5^2}\int d^5x \sqrt{-g_5} [ - 2
\Lambda + {\mathcal R} ] + \sum_{i=\pm} \frac{1}{\kappa^2_5}
\int_i d^4x \sqrt{-\bar{g}^i} (-T_i + 2 K_i) \label{RSaction}
\end{equation}
where ${\mathcal R}$ is the Ricci scalar, $\kappa_5^2$ the
gravitational constant, and $\bar{g}^i_{\mu \nu}$ denotes the
induced metric on the $i$th brane. We have also included the
Gibbons-Hawking boundary term for outgoing normal vectors. The
AdS$_5$ warped solution with 4D Poincar\'e invariance and $Z_2$
symmetry about each brane, located at constant $z$, is
\begin{equation}
\label{RSmetric} ds^2 = e^{-2 k z} \eta_{\mu\nu} dx^{\mu} dx^{\nu}
+ dz^2,
\end{equation}
requiring the well known fine-tunings
\begin{eqnarray}
\Lambda &=& -6k^2 \qquad  \qquad \;(<0)
\nonumber \\
T_+ &=& - \, T_- \; \; = \; \; 6k \; \; \;  (>0) .
\end{eqnarray}

In general, when matter is added to the branes, the physics of the
RS model cannot be derived from a 4D action since the brane is not
decoupled from the bulk, and hence the system of brane equations
is not closed \cite{Shiromizu-MS}. At {\it low energies} $E \ll
|T_i|$, the situation is different and the 4D low energy effective
action corresponding to (\ref{RSaction}) has been thoroughly
studied. In this limit, the degrees of freedom are the two brane
positions and the 4D graviton zero mode \cite{Garriga,CGR}. In the
Einstein frame, one of the two moduli, the dilaton, decouples
leaving only one physical modulus, the radion. As stressed in
\cite{nonlinear}, the resulting effective action is
non-perturbative and hence can describe the physics of strong
gravity systems such as black-holes on the brane \cite{Toby}.

Our aim is to derive a similar 4D low energy effective action when
Gauss-Bonnet (GB) gravity rather than Einstein gravity acts in the
bulk. This particular higher derivative combination is the only
one which gives equations of motion depending on the metric and
its first two derivatives: \ba S_{{\rm GB}} &=& \frac{1}{2
\kappa_5^2}\int d^5x \sqrt{-g_5} \left[- 2 \Lambda + {\mathcal R}
+ \alpha ( {\mathcal R}^2 -4 {\mathcal R}_{ab} {\mathcal
R}^{ab}+{\mathcal R}_{abcd} {\mathcal R}^{abcd}) \right]
 \nn \\
&& + \sum_{i=\pm} \frac{1}{\kappa^2_5} \int_i d^4x
\sqrt{-\bar{g}^i} (-T_i + 2{\mathcal L}_{\rm boundary}^i),
\label{GBaction} \ea where the boundary term are given in
\cite{scal6}. The coupling constant $\alpha$ has mass dimension
$-2$, and when interpreted as the string slope in a derivative
expansion, $\alpha > 0$.  Action (\ref{GBaction}) has a solution
of the same form as (\ref{RSmetric}), but now with
corrections\footnote{Note that due to an improper brane delta
function regularization, the corresponding relations given in
\cite{GB1,GB2} have an incorrect coefficient.} linear in $\alpha$
\cite{stat}
\begin{eqnarray}
\label{et la bete} \Lambda &=& -6k^2(1-2\alpha k^2)
\nonumber \\
T_+ &=& - \, T_- \; = \; 6k \left(1-\frac{4}{3}\alpha k^2 \right).
\end{eqnarray}

Static brane worlds with Gauss-Bonnet gravity have been
intensively studied \cite{stat} while time dependent solutions
have also been considered in \cite{GB1,cosm,cosm15}.  The addition
of a bulk scalar field has been  investigated in
\cite{scal6,scal9,scal}.  However, the effective brane gravity in
a system consisting of \emph{two} Minkowski branes and
Gauss-Bonnet gravity in the bulk has not yet been studied: it is
the aim of this paper.

As opposed to (\ref{RSaction}), the action (\ref{GBaction}) is not
a suitable starting point to derive a low energy effective action
when GB gravity acts in the bulk. One reason is that in contrast
with the RS model, the AdS solution (\ref{et la bete}) is
unstable: the spin 2 fluctuations contain a tachyonic mode which
is localised around the negative tension brane \cite{CD}. This
instability is a generic problem of any GB system containing a
negative tension brane. Clearly in order for the effective action
to make any sense, this mode must be `removed'. Here we follow the
procedure analysed in \cite{CD} and add induced gravity terms to
the brane so that the 5D action we consider is \be S_{\rm total
GB} = S_{{\rm GB}} + S_{{\rm ind}} \label{totalmess}\ee where
\begin{equation} S_{{\rm ind}} =
\sum_{i=\pm}\frac{\beta_i}{2\kappa_5^2} \int_i d^4x
\sqrt{-\bar{g}^i}
\bar{{\mathcal R}}_i
\end{equation}
where $\bar{{\mathcal R}}_i$ is the Ricci scalar constructed from
the 4D induced metric on each brane $\bar{g}_{\mu \nu}^{i}$. The
required constraints on $\beta_i$ have been discussed in \cite{CD}
(see also (\ref{halte-aux-tachyons})). Note that warped brane
worlds with brane curvature terms have been studied before, for
instance in \cite{DGP1,DGP2}.

The outline of the paper is the following. First we recall the
linear equations of motion for GB brane worlds with two branes and
induced gravity on each brane. We  analyse the high energy regime
from the point of view of an observer on the positive tension
brane. We find that the coupling to matter on the negative tension
brane is exponentially suppressed. Gravity becomes 4D with a
Brans--Dicke coupling to the brane bending mode. At intermediate
energy, gravity becomes 5D while the brane bending mode retains
its 4D character. Finally at low energy, we find that the
effective gravity and brane bending equations are equivalent to
the field equations obtained from an effective action involving
only one scalar field, i.e.~the radion. We then consider the same
brane world model from the point of view of the moduli
approximation and show that the resulting action obtained after
integration over the fifth dimension differs from the low energy
action derived from the linear equations of motion.

\section{Low energy action and linear equations of motion}

\subsection{Propagator}

Following \cite{Garriga,CGR}, we first give the equations of
motion for perturbations about the background solution given in
(\ref{RSmetric}) and (\ref{et la bete}). Starting from a general
gauge for the metric with the two branes located at constant
unperturbed positions $\xi_0^\pm$, we then impose the GN gauge
$h_{\mu 5}=h_{55}=0$, so that the perturbed metric takes the form
\begin{equation}
ds^2= (a^2(z)\eta_{\mu\nu} + h_{\mu\nu}) + dz^2 \label{RSgauge},
\end{equation}
where
\begin{equation}
a(z) = e^{-k z}.
\end{equation}
In addition we furthermore impose the transverse-traceless gauge
condition
\begin{equation}
{h}\equiv \eta^{\mu\nu}h_{\mu\nu} = 0 = \partial_{\mu} {h}^{\mu
\nu}
\end{equation}
so that the branes are no longer straight but located at perturbed
positions
\begin{equation}
z^\pm(x) =\xi_0^\pm+\xi^\pm(x) .
\end{equation}
 Note that throughout the following,
4D indices are raised with the flat metric $\eta_{\mu \nu}$.
Furthermore it will be useful to introduce
\begin{equation}
\gamma_{\mu \nu} = a^{-2}(z) h_{\mu \nu}.
\end{equation}

The perturbed bulk Einstein equations now take the form \cite{scal9}
\begin{equation}
(1-4\alpha k^2)(\partial^2_z-4k\partial_z+a^{-2}\Box^{(4)})
\gamma_{\mu\nu}=0 \label{bulk-eqn}
\end{equation}
where the GB term acts as an overall multiplicative constant.
Thus, as long as $4\alpha k^2 \neq 1$, the solution of
(\ref{bulk-eqn}) is just as in the RS model: in momentum space,
where $\Box^{(4)}\gamma_{\mu\nu}=-p^2 \gamma_{\mu\nu}$, it is
given by
\begin{equation}
{\gamma}_{\mu\nu}(p,z)= - \frac{(k y)^{2}}{p^2} \Big(A_{\mu\nu}(p)
J_2(y) +B_{\mu\nu}(p) Y_2(y)\Big). \label{soln}
\end{equation}
Here
\begin{equation}
y = \frac{\sqrt{-p^2}}{ka(z)} \label{ydef}
\end{equation}
is the conformal variable rescaled by $\sqrt{-p^2}$ and $J_2$,
$Y_2$ are the Bessel functions of the first and second kind.  The
$p$-dependent functions $A_{\mu\nu}$ and $B_{\mu\nu}$ are
determined by the boundary conditions for the gravitational
perturbation which, in this gauge, are given by \cite{scal9}
\begin{equation}
\left. \partial_z \gamma_{\mu\nu}(p,z)\right|_{\pm} - p^2
\ell_{\pm} a_\pm^{-1} \left. \gamma_{\mu\nu}(p,z) \right|_\pm= \mp
\kappa_5^2 a_\pm^{-2}\Sigma_{\mu\nu}^\pm(p). \label{bd}
\end{equation}
Here
\be
a_{\pm} = a(\xi^\pm_0)
\ee
are the scale factors at the unperturbed brane positions, and the
length scales $\ell_{\pm}$ are given by
\begin{equation}
\ell_\pm = \frac{1}{k a_\pm} \left( \frac{\pm
\beta_{\pm} k+8\alpha k^2}{2 (1-4\alpha k^2) }
\right). \label{elldef}
\end{equation}
These scales, which will play an important r\^ole later, vanish in
the RS limit but more generally can be either positive or
negative. Note that in (\ref{bd}) we have added matter with
stress-energy tensor $T_{\mu \nu}^{\pm}$ to each brane so that the
source term is
\begin{equation}
\Sigma_{\mu\nu}^{\pm}= \frac{1}{(1-4\alpha k^2)} \left[
\left(T^\pm_{\mu\nu} -\frac{1}{3} T^\pm \eta_{\mu\nu}\right) \mp
2\kappa_5^{-2} w_{\pm}
\partial_\mu\partial_\nu \xi^\pm \right] \label{sigma}
\end{equation}
where we have defined
\be
w_{\pm} = (1 \pm \beta_\pm k+4\alpha k^2).
\ee
The stress-energy tensors are defined with respect to the induced
metrics:
\begin{equation}
T^\pm_{\mu\nu}\equiv -\frac{2}{\sqrt{-\bar{g}_{\pm}}}\frac{\delta
{\cal L}^\pm_{{\rm matter}}}{\delta \bar{g}^{\mu\nu}_{\pm}}.
\end{equation}
Finally, the relative signs in (\ref{bd}) arise from the change of
orientation on the second brane compared to the first brane, and
these equations generalise those of \cite{Garriga} to GB gravity.
From (\ref{bd}) and $\gamma=0$, it follows that $\Sigma^{\pm}=0$
and hence
\begin{equation}
\Box^{(4)} \xi^{\pm} =\mp \frac{\kappa_5^2}{6w_{\pm}} T^{\pm}
\label{identify scalar}.
\end{equation}
On substituting the solution (\ref{soln}) into (\ref{bd}), the
boundary conditions become
\be
ky_\pm  \left\{ A_{\mu \nu}(p) \tilde{J}_\pm(p)+ B_{\mu
\nu}(p)\tilde{Y}_\pm(p) \right\} = \mp \kappa_5^2
\Sigma^{\pm}_{\mu \nu}(p) \label{bda}
\ee
where from (\ref{ydef})
\be
y_\pm = y_\pm(p) = \frac{\sqrt{-p^2}}{k a_\pm}
\ee
and
\ba
\tilde{J}_\pm(p) &\equiv&  J_1(y_\pm) + (k\ell_\pm) a_\pm  y_\pm
J_2(y_\pm), \label{tildeJdef}
\\
\tilde{Y}_\pm(p) &\equiv&  Y_1(y_\pm) + (k\ell_\pm) a_\pm y_\pm
Y_2(y_\pm). \label{tildeYdef}
\ea

When there is no matter on both branes, $\Sigma_{\mu
\nu}^{\pm}=0$, a first solution of (\ref{bda}) is when $y_\pm=0$
so that $ p^2=0$
--- the zero mode corresponding to the massless graviton. The
other solutions are obtained when the relevant determinant of
(\ref{bda}) vanishes:
\be
{\rm Det}(p) \equiv \tilde{J}_-(p) \tilde{Y}_+(p) -
\tilde{J}_+(p)\tilde{Y}_-(p) = 0.
 \label{det}
\ee
As discussed in \cite{CD}, for $\alpha \neq 0$ and $\beta_\pm=0$,
equation (\ref{det}) has solutions when $y$ is imaginary, and
these tachyonic modes with $p^2>0$ are non-perturbative in
$\alpha$. However, for non-zero induced gravity terms $\beta_\pm$
they can be prevented provided \cite{CD}
\begin{equation}
\ell_+ \ell_-< 0. \label{halte-aux-tachyons}
\end{equation}
For real $y_\pm$, equation (\ref{det}) yields the
Kaluza--Klein  tower.

We now assume that (\ref{halte-aux-tachyons}) holds and solve the
linear equation in the presence of matter on both branes. From the
boundary conditions (\ref{bda}) we find (away from the locus ${\rm
Det}(p)=0$ which corresponds to a discrete spectrum in $p^2$)
\ba
A_{\mu\nu}(p) &=& \frac{\kappa_5^2}{k} \frac{1}{{\rm Det}(p)}
\left( \frac{\Sigma^+_{\mu \nu}(p) \tilde{Y}_-(p)}{y_+} +
\frac{\Sigma^-_{\mu \nu}(p) \tilde{Y}_+(p)}{y_-} \right),
\label{Aexact}
\\
B_{\mu\nu}(p) &=& - \frac{\kappa_5^2}{k} \frac{1}{{\rm Det}(p)}
\left( \frac{\Sigma^+_{\mu \nu}(p) \tilde{J}_-(p)}{y_+} +
\frac{\Sigma^-_{\mu \nu}(p) \tilde{J}_+(p)}{y_-} \right).
\label{Bexact}
\ea
Thus, from (\ref{soln}), the general solution for $\gamma_{\mu
\nu}$ is
\begin{equation}
h_{\mu\nu}(x,z)=a^2(z)\gamma_{\mu\nu}(x,z)=\int
d^4x'\Big(\Delta^+(x,x',z) \Sigma^+_{\mu\nu}(x')
+\Delta^-(x,x',z)\Sigma^-_{\mu\nu}(x')\Big)
\label{TT metric}
\end{equation}
where the propagators are given by
\ba
\Delta^{\pm}(x,x',z) &\equiv& \int \frac{d^4p}{(2\pi)^4} e^{ip.(x'-x)}
\Delta^{\pm}(p,z)
\nonumber
\\
&=& \int \frac{d^4p}{(2\pi)^4} e^{ip.(x'-x)} \frac{\kappa_5^2
a_\pm}{\sqrt{-p^2}} \left( \frac{\tilde Y_\mp(p)J_2(y)-\tilde
J_\mp(p)Y_2(y)}{{\rm Det}(p)} \right) \label{propagator}\ea with
$y=y(p,z)$ given in (\ref{ydef}).

Finally, from (\ref{TT metric}), the perturbed metric on each
brane can be calculated.  For the positive (resp.~negative)
tension brane, we transform to a GN coordinate system $\tilde
x^a=x^a - \xi^a$ giving a straight brane located at $\tilde
z=\xi^\pm_0$ as well as $\tilde h_{\mu 5}=\tilde h_{55}=0$. After
a 4D gauge transformation \cite{Garriga,CGR}, the perturbed metric
on each brane is then $\tilde h_{\mu\nu}(x,\tilde z=\xi^\pm_0)$
with
\begin{eqnarray}
\tilde h_{\mu\nu}(x,\xi^\pm_0)  &=& h_{\mu\nu}-2ka^2_\pm
\eta_{\mu\nu}\xi^\pm \nonumber
\\
&=&\frac{1}{1-4\alpha k^2} \left\{ \int d^4x' \left(
\Delta^+(x,x',\xi^\pm_0)(T^+_{\mu\nu}-\frac{1}{3}\eta_{\mu\nu}T^+)(x')
\right. \right.
\nonumber \\
&& \qquad  \qquad \qquad + \left. \left.
\Delta^-(x,x',\xi^\pm_0)(T^-_{\mu\nu}-\frac{1}{3}\eta_{\mu\nu}T^-)(x')
\right) \right\} \pm \frac{k\kappa_5^2 a_\pm^2}{3 w_\pm}
\frac{1}{\Box^{(4)}}T^\pm \eta_{\mu\nu}
\nonumber \\
& &\qquad \qquad
\end{eqnarray}
This expression together with (\ref{propagator}) and
(\ref{identify scalar}) captures the physics of the Gauss--Bonnet
brane world models with induced gravity on the branes. Notice that
the perturbation $\tilde{h}_{\mu\nu}(x,z)$ depends on the sources
on both branes. In particular the brane positions play an
important r\^ole in the dynamics of the system. At low energy, we
will show that there is only one  effective scalar degree of
freedom. Before considering the low energy action reproducing the
linear equations of motion, let us concentrate on the high energy
regime.

\subsection{High energy limit}

At high energy, the effect of the induced brane terms is highly
relevant. In particular, we find that at very high energy gravity
propagates in 4D while its behaviour is 5D in an intermediate
range.

Consider first the positive tension brane and set
$T_{\mu\nu}^-=\xi^-=0$. In order to evaluate the propagators on
the positive tension brane it is convenient to work in Euclidean
space and define $q=-i \sqrt{-p^2}$ with $q$ real. Notice that
this also corresponds to space-like momenta $p^2
> 0$ as relevant when computing the static potential between point
sources.  In the high energy limit
$|y_\pm|=|q|/(ka_\pm)\gg 1$, we obtain the propagator
\begin{equation}
\Delta^+(q,\xi_0^+) \approx \frac{\kappa_5^2 a_+}{ q} \left(
\frac{1}{q \ell_++1} \right),
\end{equation}
from which two different energy regimes appear.
\begin{itemize}
\item
At large momenta or small distances, $q^{-1} \ll \vert
\ell_+\vert$ the propagator $\propto q^{-2}$ leading to
\begin{eqnarray}
\frac{1}{a_+^2}\tilde h_{\mu\nu}(q,\xi^\pm_0)=\frac{1}{q^2}\frac{2k\kappa_5^2}{\beta_+k+8\alpha k^2} \left[T^+_{\mu\nu}-\frac{1}{2}\eta_{\mu\nu}T^++\frac{1-4\alpha k^2}{6w_+}\eta_{\mu\nu}T^+
 \right]
\nonumber \\
\equiv \frac{1}{q^2}\frac{2\kappa_4^2}{\Phi_0}\left[T^+_{\mu\nu}-\frac{1}{2}\eta_{\mu\nu}T^++\frac{1}{2(3+2\omega(\Phi_0))}\eta_{\mu\nu}T^+
 \right]
\end{eqnarray}
We consider $\frac{1}{a_+^2}\tilde h_{\mu\nu}$ as the gravitational perturbation associated to a minkowski background $\eta_{\mu\nu}$. In this limit, the interaction with matter mediated by gravity is
a \emph{four-dimensional} tensor-scalar theory which is given in a
Brans-Dicke parametrisation \cite{Will} by a background
Brans-Dicke parameter
\begin{equation}
\omega(\Phi_0)=\frac{3}{2}\frac{\beta_+ k+ 8\alpha k^2}{1-4\alpha k^2}
\end{equation}
where the background Brans-Dicke field is
\begin{equation}
\frac{\Phi_0}{\kappa_4^2} = \frac{\beta_+ k+ 8\alpha k^2}{k\kappa_5^2}
\end{equation}
and its fluctuation
\begin{equation}
\frac{\delta \Phi}{\Phi_0}= -2k \frac{1-4\alpha k^2}{\beta_+ k+8\alpha k^2}\xi^+
\end{equation}
It coincides with the results of \cite{scal9} and the Minkowski
limit in \cite{cosm15}. This should be contrasted to the RS model
in which $\ell_+=0$ and where gravity is always five-dimensional
at short distance.
\item If $k \vert \ell_+\vert  a_+ \ll 1$, there is an intermediate high-energy
regime in which $\frac{1}{ka_+}\gg q^{-1} \gg \vert \ell_+\vert $.
In this case the propagator $\propto q^{-1}$ leading to an
effective gravity given by
\begin{equation}
\tilde h_{\mu\nu}(q,\xi^+_0)=\frac{\kappa_5^2a_+}{q(1-4\alpha
k^2)}(T^+_{\mu\nu}-\frac{1}{3}\eta_{\mu\nu}T^+)-\frac{\kappa_5^2 k
a_+^2 }{3w_+ q^2 }T^+\eta_{\mu\nu}.
\end{equation}
The $1/q$ momentum dependence associated with the $1/3$ trace
factor instead of $1/2$ means that there is 5D propagation of a
combination of a 4D-tensor and a 4D-scalar mode. The term $1/q^2$
term corresponds to the 4D propagation of a 4D-scalar mode. Again
note that in the RS model one is always in this regime at high
energy.

\end{itemize}

So far we have not taken into account the presence of a second brane.
In fact one finds
\begin{equation}
\Delta^-(\xi_0^+) \approx 2 \kappa_5^2 \sqrt{a_+ a_-} e^{
\frac{q}{k}\left(\frac{1}{a_+}-\frac{1}{a_-}\right)} \left[
\frac{1}{q} \frac{1}{(1+ \ell_+ q)} \frac{1}{(1- \ell_- q)}
\right].
\end{equation}
Notice that the propagator from the negative tension brane to the
positive tension brane is exponentially suppressed, i.e. no
gravitational effect is transmitted from one brane to another at
high energy. Hence, at high energy, the two brane system behaves
like a single brane system with no influence from the second
brane.

In the following section we show that in the low energy limit, gravity
is always 4D.

\subsection{Low energy limit}
Here we are interested in determining the dynamics and the number of degrees of
freedom in the low energy limit, $y_\pm \ll 1$.  In that limit,
equations (\ref{Aexact}) and (\ref{Bexact}) reduce to
 \begin{eqnarray}
 A_{\mu\nu} &\approx & -\frac{2k\kappa_5^2}{p^2}a_-^2
 \frac{w_-\Sigma^+_{\mu\nu}
 +w_+\Sigma_{\mu\nu}^-}{w_+ -\frac{a_-^2}{a_+^2}w_-}
\\
 B_{\mu\nu} &\approx& (1-4\alpha k^2)\frac{\pi\kappa_5^2}{2k}
 \frac{\Sigma^+_{\mu\nu}
 +\frac{a_-^2}{a_+^2}\Sigma_{\mu\nu}^-}{w_+
 -\frac{a_-^2}{a_+^2}w_-}
 \end{eqnarray}
while we have
\begin{eqnarray} \Box^{(4)}
\gamma_{\mu\nu}& \approx &-\frac{4k^2}{\pi} B_{\mu\nu}
\\
& \approx  &-\frac{1}{a_+^2w_+ -a_-^2w_-} \Big [2 \kappa_4^2
(a_+^2 T^+_{\mu\nu}+ a_-^2 T_{\mu\nu}^-)
\nonumber \\
&& \; \; \; \; -4k a_+^2w_+
(\partial_{\mu}\partial_{\nu}-\eta_{\mu\nu}\Box^{(4)})\xi^+ +4k
a_-^2w_-
(\partial_{\mu}\partial_{\nu}-\eta_{\mu\nu}\Box^{(4)})\xi^-
\Big ]
 \label{identify tensor}
\end{eqnarray}
where $\kappa_4^2=k\kappa_5^2$ and we have used (\ref{identify
scalar}).

These equations have  the structure of the equations of motion
from a low energy effective action involving tensor gravity $
\gamma_{\mu\nu}(x)$ and two scalar fields $\xi^{\pm}(x)$. They can
be reproduced by a quadratic action, expanding
\begin{eqnarray}
&\frac{1}{2\kappa_4^2} \int d^4 x
\sqrt{-g}\Big([F_+(\xi^+)-F_-(\xi^-)] {\cal R}
-B_+(\xi^+)(\partial\xi^+)^2 -B_-(\xi^-) (\partial \xi^-)^2\Big)&
\nonumber \\
&+S^+_{matter}(A_+(\xi^+)g_{\mu\nu})+
S^-_{matter}(A_-(\xi^-)g_{\mu\nu})&
\end{eqnarray}
to second order around $g_{\mu\nu}=\eta_{\mu\nu}$ , $\xi^\pm=0$ .
Here matter on each brane is minimally coupled to the indicated
metric. We find that one can identify
\begin{equation}
F_\pm= \pm w_\pm a^2(\xi^\pm_0+\xi^\pm)
\end{equation}
and the sigma model coefficients
\begin{equation}
 B_\pm=\mp 6 k^2w_\pm a^2(\xi^\pm_0+\xi^\pm)
\end{equation}
The coupling functions to matter are given by
\begin{equation}
A_\pm= a^2(\xi^\pm_0+\xi^\pm)
\end{equation}
implying that matter couples to the induced metric on each brane.
Notice that in the $\alpha\to 0$ and $\beta_{\pm}\to 0$ limits,
one obtains the scalar-tensor theory corresponding to the
Randall-Sundrum case.

A quick glance at the action that we have just derived seems to
indicate that there are two scalar degrees of freedom while there
is only one effective scalar degree of freedom in the R--S case.
To determine  the structure of the effective action, it is
convenient to go to the Einstein frame where the Planck mass is
fixed. In the following we will assume that
\begin{equation}
w_+ a^2_+ > w_- a^2_-
\end{equation}
guaranteeing that the squared effective Planck mass is positive in
the brane frame (and thus in all frames so that the graviton is
not a ghost). The corresponding Einstein frame action is the quadratic
expansion, around $g_{\mu\nu}=(F_+^0-F_-^0)\eta_{\mu\nu}$ and $\xi^\pm=0$, of
\begin{equation}
S_{EF}=\frac{1}{2\kappa_4^2}\int d^4x \sqrt{-g}\Big({\cal
R}-\sigma_{ij}\partial \xi^i \partial \xi^j\Big) + S^+_{\rm
mat}(\frac{A_+}{F_+-F_-}g_{\mu\nu}) + S^-_{\rm
mat}(\frac{A_-}{F_+-F_-}g_{\mu\nu})
\end{equation}
with
\begin{equation}
\sigma_{ij}=\left(
\begin{tabular}{cc}
$\frac{3}{2}(\frac{F'_+}{F_+-F_-})^2+\frac{B_+}{F_+-F_-}$& $-\frac{3}{2}\frac{F_+' F_-'}{(F_+-F_-)^2}$\\
$-\frac{3}{2}\frac{F_+' F_-'}{(F_+-F_-)^2}$&
$\frac{3}{2}(\frac{F_-'}{F_+-F_-})^2+\frac{B_-}{F_+-F_-}$
\end{tabular}
\right)
\end{equation}
and $i,j=1..2=+,-$. This sigma model matrix simplifies drastically
in our case and takes the form
\begin{equation}
\sigma_{ij} =\frac{6k^2 F_+F_-}{(F_+-F_-)^2} \left(
\begin{tabular}{cc}
1 & -1 \\
-1 & 1
\end{tabular}
\right)
\end{equation}
It is easy to see that this matrix has a zero eigenvalue leading
to the presence of
 only one physical scalar degree of freedom, the radion
$r=R+\xi^- -\xi^+$ where $R=\xi_0^--\xi_0^+$ is the unperturbed
 interbrane distance.
Therefore the action is the quadratic expansion of
\begin{eqnarray}
S_{EF}=\frac{1}{2\kappa_4^2}\int d^4x \sqrt{-g}\Big({\cal
R}-\frac{6k^2 w_- a^2(r)}{w_+(1 -\frac{w_-}{w_+}a^2(r))^2}(\partial
r )^2\Big)
+ S^+_{\rm mat}\Big(\frac{g_{\mu\nu}}{w_+-a^2(r)w_-}\Big)
\nonumber \\
+S^-_{\rm mat}\Big(\frac{g_{\mu\nu}}{a^{-2}(r)w_+ -w_-}\Big)
\end{eqnarray}
Requiring that the radion is not a ghost implies that
\begin{equation}
w_+ w_-> 0
\end{equation}
When the graviton and the radion are not ghosts, the low energy
effective action provides useful information on the Gauss--Bonnet
brane world at low energy.

Let us assume that $\vert w_-\vert \le \vert w_+\vert$ and define
\begin{equation}
\sqrt{\frac{w_-}{w_+}}e^{-kr}=\tanh \rho
\end{equation}
The effective action becomes now (the quadratic expansion of)
\begin{equation}
S_{EF}=\frac{1}{2\kappa_4^2}\int d^4x \sqrt{-g}\Big({\cal
R}-6(\partial
\rho)^2\Big)+ S^+_{\rm mat}\Big(\frac{\cosh^2\rho }{w_+}g_{\mu\nu}\Big)
+S^-_{\rm mat}\Big(\frac{\sinh^2 \rho}{w_-}g_{\mu\nu}\Big)
\end{equation}
Notice that the only difference with the RS effective action
resides in the prefactors $w_\pm$ in the coupling of the radion to
matter. When these prefactors are equal to unity, the effective
action is the RS one as derived within the moduli space
approximation. We will compare the effective action obtained from
the linear equations of motion and the moduli space approximation
in the following section.

The coupling to gravity has to be such that the presence of a
massless degree of freedom does not modify gravity. To carry out
this analysis, it is convenient to use another form of the action.
The action can be put in the Brans-Dicke form using the metric on
each brane as the gravitational field. For the positive tension
brane matter, the action becomes (the quadratic expansion of)
\begin{equation}
S^+_{BD}=\frac{w_+}{4 k\kappa_5^2}\int d^4 x \sqrt{-g}\Big(\Psi
{\cal R}-\frac{\omega_+(\Psi)}{\Psi}(\partial \Psi)^2\Big)+
S^+_{\rm mat}(g_{\mu\nu})+S^-_{\rm mat}(A_-^{BD}(\Psi)g_{\mu\nu})
\label{BD}
\end{equation}
where the Brans--Dicke field is
\begin{equation}
\Psi = 1-\frac{w_-}{w_+}e^{-2k r}
\end{equation}
with a Brans-Dicke parameter
\begin{equation}
\omega_+(\Psi)=\frac{3}{2}\frac{\Psi}{1-\Psi}
\end{equation}
and a coupling to matter of the second brane
\begin{equation}
A^{BD}_-(\Psi)=\frac{w_+}{w_-}(1-\Psi)
\end{equation}
 Notice that the Brans--Dicke parameter can be arbitrarily large
when the branes are far apart. Hence ordinary matter can be
located on the positive tension brane. This coincides with the
usual R--S result.

Similarly for the negative tension brane this is the second order expansion of
\begin{equation}
S^-_{BD}=\frac{w_-}{4k\kappa_5^2}\int d^4 x \sqrt{-g}\Big(\Phi
{\cal R}-\frac{\omega_-(\Phi)}{\Phi}(\partial \Phi)^2\Big)+
S^+_{\rm mat}(A^{BD}_+(\Phi)g_{\mu\nu})+S^-_{\rm mat}(g_{\mu\nu})
\end{equation}
with a Brans-Dicke parameter
\begin{equation}
\omega_-(\Phi)=-\frac{3}{2} \frac{\Phi}{1+ \Phi}
\end{equation}
and a coupling to matter
\begin{equation}
A^{BD}_+(\Phi)=\frac{w_-}{w_+}(1+\Phi)
\end{equation}
where the Brans--Dicke field is
\begin{equation}
\Phi =e^{2 k r}\frac{w_+}{w_-} - 1
\end{equation}
Notice that the Brans-Dicke parameter is here negative for large
brane distances, ruling out the possibility of having ordinary
matter on the second brane.

\subsection{The projective approach}

The previous action can be retrieved using the projective approach
\cite{Shiromizu-MS,nonlinear}, in which the Einstein equations on
both branes are written in terms of the matter energy--momentum
tensors and the projected Weyl tensor $E_{\mu\nu}$. Eliminating
the projected Weyl tensor between the two brane equations leads to
the effective Einstein equation on either brane. Here we will
concentrate on the case $\alpha=0$ for simplicity. The general
case is beyond the scope of the present paper. At low energy one
can neglect the quadratic terms in the matter content of the
branes. The Einstein equation on the first brane reads
\begin{equation}
w_+G_{\mu\nu}(\bar g_{\mu\nu}^+) =k\kappa_5^2 T^{ +}_{\mu\nu}
-E_{\mu\nu}
\end{equation}
where we have indicated the dependence on the induced metric
explicitly, and the contribution in $\beta_+$ comes from the brane
curvature term.  Similarly, on the second brane,
\begin{equation}
w_-G_{\mu\nu}(\bar g_{\mu\nu}^-) =k\kappa_5^2 T_{\mu \nu}^-
-\frac{E_{\mu\nu}}{\Omega^4}
\end{equation}
where $\Omega=\frac{a_-}{a_+}$ corresponds to the radion field.
Using $\bar g_{\mu\nu}^-= \Omega^2 \bar g_{\mu\nu}^+$ to lowest
order in a derivative expansion, one can eliminate $E_{\mu \nu}$
and obtain the Einstein equation
\begin{eqnarray}
G_{\mu\nu}(\bar g^+_{\mu\nu})= &
\frac{\kappa_4^2}{\Psi}(T^+_{\mu\nu}+
\frac{w_+}{w_-}(1-\Psi)T_{\mu\nu}^-) +
\frac{\omega(\Psi)}{\Psi^2}(D_\mu\Psi D_\nu\Psi -\frac{1}{2}
(D\Psi)^2 \bar g^+_{\mu\nu})&
\nonumber \\
&+ \frac{1}{\Psi} ( D_\mu D_\nu \Psi - D^2\Psi \bar g^+_{\mu\nu})&
\end{eqnarray}
which coincides with the Einstein equations deduced from the
effective action obtained in the previous section. Hence, the
projective approach leads to the same results as the linear
equations of motion. In the RS case, it has been shown that the
effective action can also be deduced using   the moduli space
approximation. In the following section, we examine the validity
of the moduli space approximation for Gauss--Bonnet brane worlds.

\section{The Moduli Space Approximation}

The previous action has been obtained from the linear equations of
motion (also the projective approach). In the RS case, it has also
been shown that the action can
 be obtained from the moduli approximation, a method more akin
to the ones used in string theory where one integrates over a
compact manifold in order to retrieve a
four dimensional action.

We will derive the 4D low-energy effective action in the moduli
space approximation, that is by keeping only the massless degrees
of freedom represented here by the 4D metric $g_{\mu\nu}$ plus two
real 4D scalar fields giving the brane positions in the fifth
dimension:
\begin{equation}
ds^2 = a^2(x^5) g_{\mu\nu}(x^{\lambda}) dx^{\mu} dx^{\nu} +
(dx^5)^2
\label{sub}.
\end{equation}
The low
energy approximation consists in integrating over the extra dimension and expanding the action
in derivatives of the moduli fields,
implying that we consider only slowly varying brane positions.

In this section we derive the low energy effective action for
$S_{GB}$ given in (\ref{GBaction}).  To do this, let $x^5_i =
\phi_i(x^{\mu})$ be the two brane positions. We
substitute the ansatz (\ref{sub}) into $S_{{\rm GB}}$ and
integrate over the range $\phi_+(x^{\mu}) \leq z \leq
\phi_-(x^{\mu})$. The full effective action for a general warp
factor, $S=S_{bulk}+S_{branes}$, reads
\begin{eqnarray}
S_{bulk} &=& \frac{1}{2\kappa_5^2}\int d^4x \sqrt{-g}[
\bar{{\mathcal R}} \left\{ \int
dx^5\Big(a^2-\alpha(4a'^2+8aa'')\Big)\right\}
\nonumber \\
&& \qquad \qquad  \qquad \; \; + \int dx^5\Big(-2\Lambda a^4 -8
a''a^3-12a'^2 a^2+24\alpha(a'^4+4a''a'^2a) \Big)]
\nonumber \\
\label{wow1} \\
S^+_{brane}&=&-\frac{1}{\kappa_5^2}\int d^4x \sqrt{-g}\Big[
(8\alpha a_+a_+' -\frac{\beta_+ a_+^2}{2})\bar{{\mathcal R}} +
(T_+ a_+^4 + 8 a_+'a_+^3-32\alpha a_+ a_+'^3 )
\nonumber \\
&&\qquad \qquad  \qquad \; \;  +
\left(2a_+a_+'+T_+\frac{a_+^2}{2}+8\alpha
\frac{a_+'^3}{a_+}\right)(\partial \phi_+)^2
  + (-2a_+^2+24\alpha a_+'^2)\Box
\phi_+ \Big] . \nonumber \\ \label{wow2}
\end{eqnarray}
Here $ \int dx^5 = 2\int_{\phi_+}^{\phi_-} dx^5 $ where the factor
of 2 is to take account of the $Z_2$ symmetry. In $S_{brane}^+$
the warp factor $a_+$ and its derivatives are evaluated at
$x^5=\phi_+$. A similar expression gives the brane action of the
negative tension brane.

In the case of the exponential background (\ref{RSmetric}) with
$a(x^5)=\exp(-kx^5)$, evaluation of the integrals in (\ref{wow1})
followed by the addition of (\ref{wow2}) for both branes gives
\begin{eqnarray}
S &=& \frac{1}{\kappa_5^2}\int d^4x \sqrt{-g} \left[ \frac{w_+
e^{-2k\phi_+}-w_-e^{-2k\phi_-}}{2 k}\bar{\mathcal R} \right.
\nonumber \\
&&\qquad \qquad  \qquad \; \; + \, \left(6k-\frac{1}{2}T_+ -
40\alpha k^3\right)e^{-2k\phi_+}(\partial \phi_+)^2
\nonumber \\
&& \qquad \qquad  \qquad \; \; - \, \left(6k+\frac{1}{2}T_- -
40\alpha k^3\right)e^{-2k\phi_-}(\partial \phi_-)^2
\nonumber \\
&& \qquad \qquad  \qquad \; \; - \,
\left(-3k+\frac{\Lambda}{2k}+\frac{T_+-T_-}{2}+2\alpha
k^3\right)\left(e^{-4k\phi_+}-e^{-4k\phi_-}\right)
\nonumber \\
&& \left. \qquad \qquad  \qquad \; \; - \,
\frac{T_++T_-}{2}(e^{-4k\phi_+}+e^{-4k\phi_-})\right] ..
\label{effective action}
\end{eqnarray}

It is very interesting to remark that, at first sight, the moduli
fields $\phi_i$ appear to have a non-vanishing potential in
(\ref{effective action}). However, once the relations (\ref{et la
bete}) have been used, the potential vanishes identically.  This
convenient check is in accordance with Birkhoff's theorem: any
brane position at constant $x^5$ must be solution of the effective
action for the fine-tuned tensions satisfying (\ref{et la bete}).

In order to compare this action with the effective action deduced from the linear equations of motion, it is convenient to go to the Einstein
frame.
We will assume that the two prefactors
$w_+ $ and $w_-$ are
positive. The field redefinition simplifying the Einstein-Hilbert
term  becomes
\begin{eqnarray}
e^{-k\phi_+} &=& \frac{1}{\sqrt{w_+}}e^{\sigma}\cosh(\rho)
\nonumber \\
e^{-k\phi_-} &=& \frac{1}{\sqrt{w_- }}e^{\sigma}\sinh(\rho)
\end{eqnarray}
Notice that $\rho$ is a decreasing function of the interbrane
distance whilst $\sigma$ is related to the "center of mass"
position of the two-brane system along the fifth dimension.
This yields the Einstein frame effective action
\begin{equation}
S =\frac{1}{2k\kappa_5^2}\int d^4x \sqrt{-g} \Big[ {\mathcal R} -
\gamma_{\sigma \sigma}(\partial \sigma)^2 - \gamma_{\rho\rho}
(\partial \rho)^2 -2\gamma_{\rho\sigma}(\partial \rho)(\partial
\sigma)\Big]
\end{equation}
with the normalization matrix
\begin{eqnarray}
\gamma_{\sigma \sigma}&=&\frac{96\alpha k^2+6 \beta_+ k}{1+4\alpha
k^2 +\beta_+ k}\cosh^2(\rho)  - \frac{96 \alpha k^2-6 \beta_-
k}{1+4\alpha k^2-\beta_- k}\sinh^2(\rho)
\nonumber \\
\gamma_{\rho\rho}&=& \frac{6-72\alpha k^2} {1+4\alpha k^2-\beta_-
k}\cosh^2(\rho)-\frac{6-72\alpha k^2 }{1+4\alpha k^2+\beta_+
k}\sinh^2(\rho)
\nonumber \\
\gamma_{\rho\sigma}&=&\frac{k(\beta_++\beta_-)(6-72\alpha
k^2)}{(1+4\alpha k^2+\beta_+ k)(1+4\alpha k^2-\beta_- k)}
\sinh(\rho)\cosh(\rho) .
\end{eqnarray}
This is the effective action of a Gauss--Bonnet brane  world in
the moduli approximation. As can be easily seen the sigma model
matrix is of rank two, leading to the existence of two massless
degrees of freedom in the scalar sector. This contradicts the
linear equations and therefore invalidates the moduli
approximation in the Gauss--Bonnet case. The failure of the moduli
space approximation here, and the non-equivalence with the
projective approach deserves further study. In particular, its
link with either the presence of higher derivative terms  or the
necessity of extending the moduli ansatz needs to be investigated.
This is left for future work.

\section{Conclusion}

We have analysed brane worlds with a bulk Gauss--Bonnet term and
induced brane gravity terms. We have studied the high energy and
low energy limits. In particular, we have shown that the low
energy effective action involves only one field, the radion, and
differs from the RS case. The difference with the RS case arises
in the coupling of the radion to matter and the value of the
effective Planck mass.

We have also noted that the moduli approximation fails for
Gauss--Bonnet brane worlds. Indeed it fails to reproduce the
linear equations of motion and involves a spurious scalar degree
of freedom. This means that dimensional reduction does not commute
with taking the equations of motions from the action; the correct
procedure consists in first taking the higher dimensional
equations of motion and then dimensionally reducing them. Similar
cases of non-commutativity have been described in \cite{roy} where
it is specifically due to the Gauss-Bonnet term, or in
\cite{non-comm} where it has been shown more generally that it can
arise from symmetries of the equations of motion which are not
symmetries of the action. In this context, a better understanding
of the link between the moduli approximation and the projective
approach  deserves to be further investigated and is left for
future work.

\section{Acknowledgments}
We would like to thank C.~van de Bruck, C.~Charmousis,
J.-F.~Dufaux and S.~Davis for useful discussions and comments.
Ph.~B.~and N.~C.~are partially supported by the E.U. Research
Training Network ``The Quest for Unification: Theory Confronts
Experiment", MRTN-CT-2004-503369.

\end{document}